\documentclass [10pt]{article}
\usepackage[T1]{fontenc}
\usepackage[final]{epsfig}
\usepackage{amssymb}
\usepackage{multicol}
\usepackage{graphics}
\usepackage{ntimes}
\usepackage{color}

\newcommand{\R}{\mathbb{R}}
\newcommand{\Ham}{{\mathcal H}}
\newcommand{\Z}{{\mathcal Z}}
\newcommand{\Zred}{{\mathcal Z_{N}}}
\newcommand{\IM}{{\mathtt i}}

\newcommand{\BE}{\begin{equation}}
\newcommand{\EE}{\end{equation}}

\begin{document}

\title{Equilibrium distributions in thermodynamical traffic gas}

\author{\textbf{Milan Krb\'alek}\\
Faculty of Nuclear Sciences and Physical Engineering,\\
Czech Technical University,\\  Trojanova 13, 120 00 Prague, Czech
Republic}

\maketitle

\begin{abstract}
We derive the exact formula for thermal-equilibrium spacing
distribution of one-dimensional particle gas with repulsive
potential $V(r)=r^{-\alpha}$ $(\alpha>0)$ depending on the
distance $r$ between the neighboring particles. The calculated
distribution (for $\alpha=1$) is successfully compared with the
highway-traffic clearance distributions, which provides a detailed
view of changes in microscopical structure of traffic sample
depending on traffic density. In addition to that, the observed
correspondence is a strong support of studies applying the
equilibrium statistical physics to traffic
modelling.\\

\noindent\emph{PACS numbers: 05.70-a, 05.20-y, 45.70 Vn}

\noindent\emph{\noindent Key words: vehicular traffic,
one-dimensional gas, power-law interaction potential, spacing
distribution }

\end{abstract}

%\section{Introduction}

\begin{multicols}{2}

Investigation of one-dimensional particle ensembles seems actually
to be very useful for understanding of the complex system called
\emph{vehicular traffic.} Beside the favorite cellular automata,
in the recent time new trend appears in the traffic modelling.
Application of the equilibrium statistical-physics to traffic
ensembles (queuing systems of spatially interacting particles) has
been more times discussed (see for example Ref. \cite{Helbing} and
other references in) and successfully demonstrated in the articles
\cite{HKT}, \cite{Mahnke}, and \cite{Helbing_and_Krbalek}. It has
been manifested in Ref. \cite{Mahnke} that thermodynamics approach
can be applied to such a many-particle driven system as traffic
flow, based on a microscopic description, in analogy to
equilibrium physical systems. Besides, it has been demonstrated in
Ref. \cite{Helbing_and_Krbalek} that the relevant statistical
distributions obtained from the local thermodynamical model are in
accord with those from traffic data measured on real freeways by
the induction-loop detectors. It opens up an opportunity for
finding the analytical form of probability density for clear
distance among the cars (i.e. \emph{clearance distribution} in
traffic terminology) moving in traffic stream. Up to the present
time the question about analytical form of clearance distribution
has only been a subject of speculations (see the Ref.
\cite{Neubert}, \cite{Helbing}) -- never successfully answered.\\

We aim to use one-dimensional thermodynamical particle gas to the
prediction of microscopical structure in traffic flows and
consequently make a comparison to the relevant traffic data
distributions. Justification for the approach described can be
found in Ref. \cite{HKT} and \cite{Helbing_and_Treiber} where it
is proved that equilibrium solution of certain family of the
particle gases (exposed to the heat bath with the temperature $T
\geq 0$) is a good approximation for steady-state solution of
driven many-particle systems with asymmetrical interactions. Since
the vehicular traffic is a dissipative system of active elements
(moving far from equilibrium), it is evi\-dent (see Ref.
\cite{Mahnke} and \cite{HKT}) that thermodynamics-balance approach
can be used on a mesoscopic level only (i.e. for small-sized
samples of $N$ vehicles), where the traffic density fluctuates
around the constant value and distances and velocities are
mutually uncorrelated. Then the stationary solution of relevant
Fokker-Planck equation practically coincides with thermal-balance
probability density of certain statistical gas (see Ref.
\cite{HKT}). Moreover, the possibility for using the
thermodynamical approach is supported by the fact that
distribution of velocities in traffic sample fully corresponds to
the Gaussian distribution (in each traffic-density interval) --
see Ref. \cite{Helbing}, \cite{Gauss}, and
\cite{Helbing_and_Krbalek}. Overall, it is sufficiently justified
that traffic systems can be \emph{locally} (i.e. on a mesoscopic
level) described by instruments of equilibrium statistical
physics.\\

Thus, consider $N$ identical particles (vehicles) on the circle of
the circumference $L=N.$ Let $x_i$ $(i=1\ldots N)$ denote the
circular position of the $i$th particle. Put $x_{N+1}=x_1+2\pi,$
for convenience. Now we introduce the short-ranged potential
energy
$$U \propto \sum_{i=1}^N V\left(r_i\right),$$
where $V(r_i)$ corresponds to the repulsive two-body potential
depending on the distance $r_i=|x_{i+1}-x_i|\frac{N}{2\pi}$
between the neighboring particles only. Nearest-neighbor
interaction is chosen with the respect to the realistic behavior
of car-driver in traffic sample (see Ref.
\cite{Helbing_and_Krbalek}). Besides, the potential $V(r)$ has to
be defined so that $\lim_{r\rightarrow 0_+}V(r)=\infty$ which
prevents particles passing through each other. The hamiltonian of
the described ensemble reads as
$$\Ham=\frac{1}{2}\sum_{i=1}^N (v_i-\overline{v})^2+C\sum_{i=1}^N
V\left(r_i\right),$$
with the $i$th particle velocity $v_i$ and the positive constant
$C.$ Note that $\overline{v}$ represents the mean velocity in the
ensemble. Then, the appropriate partition
function\footnote{$\sigma$ is the constant representing a
statistical variance}
\begin{equation}
\Z = \int_{\R^{2N}} \delta \left(L-\sum_{i=1}^N
r_i\right)\prod_{i=1}^N
e^{-\frac{(v_i-\overline{v})^2}{2\sigma^2}}~e^{-C\frac{V(r_i)}{\sigma^2}}
dr_idv_i \label{Partition_function01}
\end{equation}
leads us (after $2N-1$ integrations) to the simple assertion that
velocity $v$ of particles is Gaussian distributed, i.e.
$$P(v)=\frac{1}{\sqrt{2\pi} \sigma}~ e^{-\frac{(v-\overline{v})^2}{2\sigma^2}}$$
is the corresponding probability density.\\

Of larger interest, however, is the spacing distribution
$P_\beta(r)$. To calculate the exact form of $P_\beta(r)$ one can
restrict the partition function (\ref{Partition_function01}) by
$N$ velocity-integrations to the reduced form
$$\Zred(L)=\int_{\R^N} \delta \left( L-\sum_{i=1}^N r_i\right) e^{-\beta\sum_{i=1}^N V\left(r_i\right)}\,dr_1\ldots\,dr_N$$
where $\beta=C\sigma^{-2}$ is \emph{the inverse temperature}
(dimensionless) of the heat bath. Denoting $f(r)=e^{-\beta V(r)}$
the previous expression changes to
$$\Zred(L)=\int_{\R^N} \delta \left( L-\sum_{i=1}^N r_i\right) \prod_{i=1}^N f(r_i)~dr_1\ldots\,dr_N.$$
Applying the Laplace transformation (see the Ref.
\cite{Bogomolny}
for details) one can obtain
$$g_N(p) \equiv \int_0^\infty \Zred(L)~ e^{-pL}\,dL=$$
$$=\left( \int_0^\infty f(r) e^{-pr}\,dr \right)^N \equiv \left[
g(p) \right]^N.$$
Then the partition function (in the large $N$ limit) can be
computed with the help of Laplace inversion
$$\Zred(L)=\frac{1}{2\pi\IM} \int_{B-\IM\infty}^{B+\IM\infty}
g_N(p) ~ e^{Lp}~dp.$$
Its value is well estimated by the approximation in the saddle
point $B$ which is determined using the equation
$$\frac{1}{g(B)}\frac{\partial g}{\partial p}(B)=-\frac{L}{N}.$$
Thus,
\begin{equation} \Zred(L) \approx \left[ g(B) \right]^N e^{LB}.
\label{party_sum} \end{equation}
Hence the probability density for spacing $r_1$ between the
particles $1$ and $2$ can be then reduced to the form
$$P(r_1)=\frac{\Z_{N-1}(L-r_1)}{\Zred(L)} f(r_1).$$
Supposing $N \gg 1$ and using equation (\ref{party_sum}) we obtain
$$P(r_1)=\frac{1}{g(B)}~ f(r_1) ~ e^{-Br_1},$$
which leads (after applying the same procedure for every pair of
successive particles) to the distribution function for spacing $r$
between arbitrary couple of neighboring particles
\begin{equation}
P_\beta(r)=A ~ e^{-\beta V(r)} ~ e^{-Br}\hspace{0.5cm} (r\geq 0).
\label{LS}
\end{equation}
Note that constant $A$ assures the normalization $\int_0^\infty
P_\beta(r)~dr=1.$ Furthermore, returning to the original choice
$L=N,$ the mean spacing is
\begin{equation}
\langle r \rangle \equiv \int_0^\infty r P_\beta(r)~dr=1.
\label{Scaling}
\end{equation}
%
%\section{Normalization of the spacing distribution}
%
Two above conditions can be understood as equation system for
unknown normalization constants $A,B.$\\

Let us to proceed to the special variants of the gas studied.
Firstly, we draw our attention to the Coulomb gas with the
logarithmic potential
$$V(r):=-\ln(r) \hspace{1cm} (r>0).$$
Such a gas (usually called Dyson's gas, for example in Ref.
\cite{Scharf}) is frequently used in the many branches of physics
(including the traffic research in Ref. \cite{Wagner}) and the
corresponding spacing distribution reads as (see Ref.
\cite{Bogomolny})
$$P_\beta(r)=\frac{(\beta+1)^{\beta+1}}{\Gamma(\beta+1)}~r^\beta
e^{-(\beta+1)r},$$
where $\Gamma(\xi)$ is gamma function. Of larger physical
interest, as demonstrated in Ref. \cite{HKT} and
\cite{Helbing_and_Krbalek}, seem actually to be the potentials
$$V_\alpha(r):=r^{-\alpha}\hspace{1cm} (r>0),$$
for $\alpha>0.$ The aim of the following computational procedure
is to normalize the distribution
\begin{equation}
P_\beta(r)=Ae^{-\frac{\beta}{r^\alpha}}e^{-Br}. \label{spacing
distribution}
\end{equation}
Consider now the favorable choice $\alpha=1,$ for which the
normalization integrals are exactly expressed as
\begin{eqnarray}
\int_0^\infty e^{-\frac{\beta}{r}} e^{-Br} ~dr= 2
\sqrt{\frac{\beta}{B}}~ \mathcal{K}_1\left(2\sqrt{\beta B} \right)
\label{kabel} \\ \int_0^\infty
re^{-\frac{\beta}{r}}e^{-Br}~dr=2\frac{\beta}{B}~\mathcal{K}_2\left(2\sqrt{\beta
B} \right), \label{label}
\end{eqnarray}
where ${\mathcal{K}}_\lambda$ is the Mac-Donald's function
(modified Bessel's function of the second kind) of order
$\lambda,$ having for $\lambda=1$ and $\lambda=2$ an approximate
expression
$$\mathcal{K}_\lambda(y) = \sqrt{\frac{\pi}{2}}e^{-y}
\left(y^{-1/2} + \frac{3}{8} 5^{\lambda-1} y^{-3/2} +
\mathcal{O}\left(y^{-5/2}\right)\right).$$
Applying the equations (\ref{kabel}) and (\ref{label}) to the
normalization integrals one can determine the exact values of the
constants $A$ and $B$. Both of them can be, after applying
Taylor's expansion procedure, very well estimated by the
approximations
\BE B\approx \beta  + \frac{3-e^{-\sqrt{\beta}}}{2},
\label{main_approximation} \EE
and
\BE A \approx
\frac{\sqrt{2\beta+3-e^{-\sqrt{\beta}}}}{\sqrt{8\beta}~\mathcal{K}_1\left(\sqrt{4\beta^2+6\beta-2\beta
e^{-\sqrt{\beta}}}\right)}. \label{A_approximation} \EE

\begin{center}
\scalebox{.4}{\includegraphics{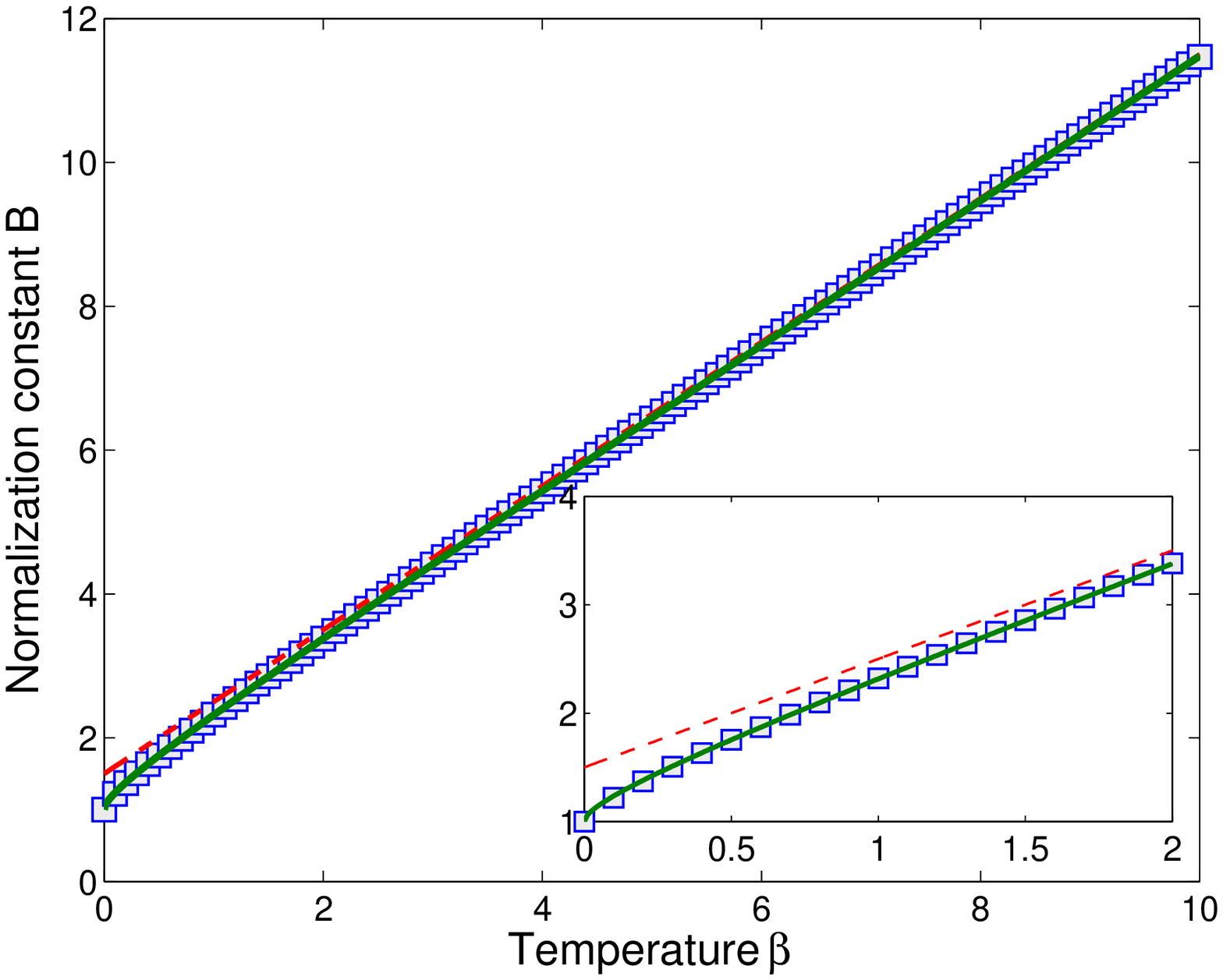}} \footnotesize

\textbf{Figure 1. Normalization constant $B$ depending on the
inverse temperature $\beta.$}\\

Squares represent the exact value of $B$ obtained from numerical
computations. The dashed and solid curves display the large
$\beta$ approximation (\ref{extim_B}) and full approximation
(\ref{main_approximation}), respectively. The behavior close to
the origin is magnified on the inset.

\end{center}

Finally, we investigate the distribution (\ref{spacing
distribution}) for general $\alpha>0.$ Although in this case the
normalization integrals are not trivially solvable, the scaling
(\ref{Scaling}) leads us to the simple approximate formula
\begin{equation} B \approx \alpha\beta+1 + \frac{\alpha}{2}\hspace{1cm}
\left(\beta\gg 1\right).\label{extim_B} \end{equation}
The large $\beta$ estimation $r^{-\alpha} \approx 1 - \alpha +
\alpha r^{-1},$ which holds true for values $r$ around mean
distance $r\approx 1,$ provides the asymptotical formula for
normalization constant $A:$
\begin{equation} A \approx \frac{1}{2}
\sqrt{1+\frac{1}{2\beta}+\frac{1}{\alpha \beta}}
\frac{e^{\beta(1-\alpha)}}{\mathcal{K}_1\left(\sqrt{2\alpha\beta
(2\alpha\beta+\alpha+2)}\right)}. \label{extim_A} \end{equation}

\begin{center}
\scalebox{.4}{\includegraphics{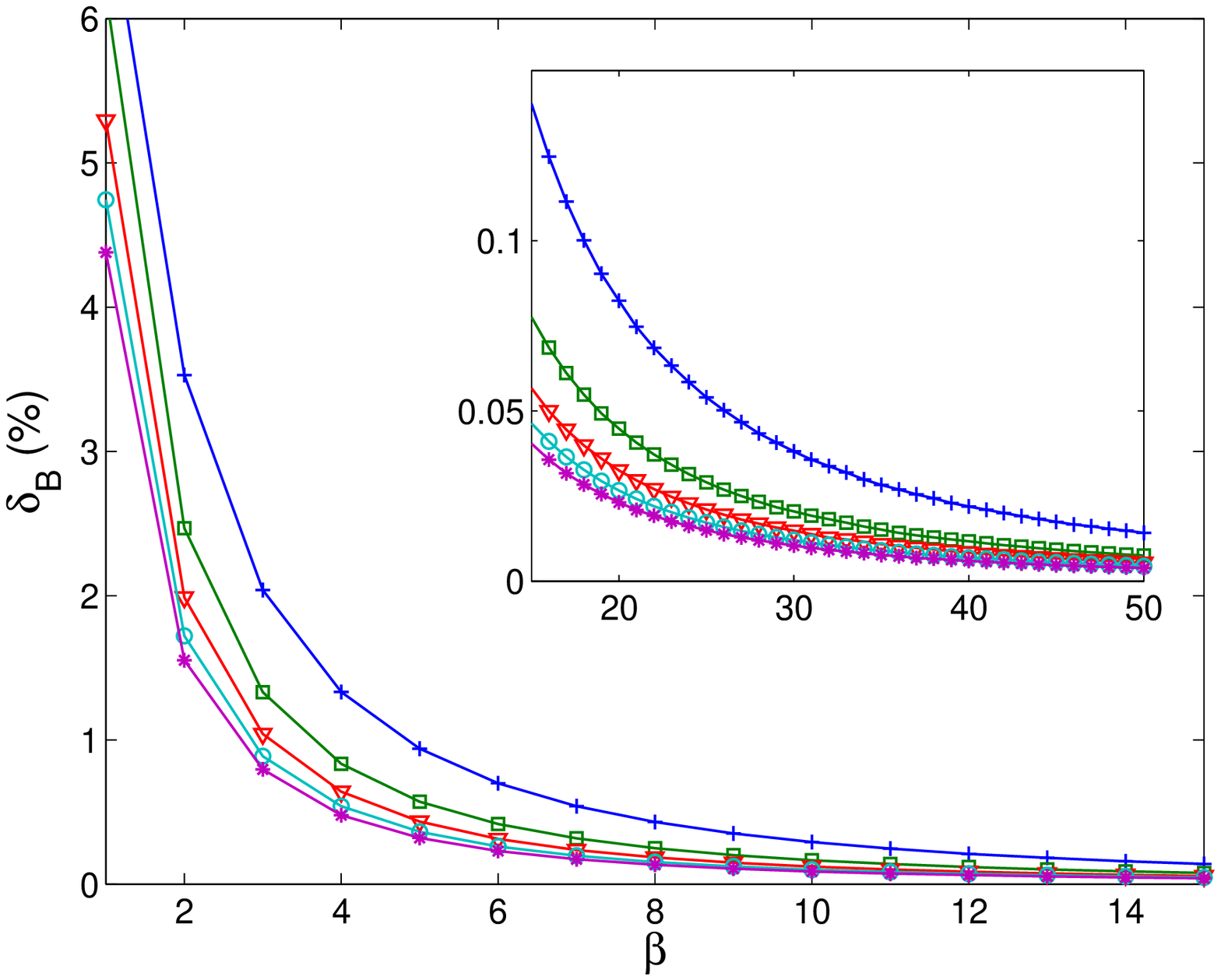}} \footnotesize

\textbf{Figure 2. Relative deviation in the approximate value of
the normalization constant $B$ as function of parameter $\beta.$}\\

We display the deviation (\ref{disc_B}) between the numerical
value $B_{ex}$ and the value $B_{est}$ obtained from the large
$\beta$ approximation (\ref{extim_B}). The plus signs, squares,
triangles, circles and stars correspond to the parameters
$\alpha=1,2,3,4,5,$ respectively. The tails of the curves are
magnified on the inset.

\end{center}

For practical applications it seems to be useful to detect the
critical inverse temperature $\beta_{crit}$ under which the
re\-la\-ti\-ve deviation between the exact ($ex$) and estimated
($est$) values of the constant A (or B)
\begin{equation}
\delta_{\log(A)}:=\frac{|\log(A_{ex})-\log(A_{est})|}{\log(A_{ex})}
\label{disc_A} \end{equation}
\begin{equation} \delta_B:=\frac{|B_{ex}-B_{est}|}{B_{ex}} \label{disc_B} \end{equation}
are larger then the fixed acceptable deviation $\delta.$  For
these purposes we plot the functional dependence
$\delta_B=\delta_B(\beta)$  and
$\delta_{\log(A)}=\delta_{\log(A)}(\beta)$ in the Fig. 2 and Fig.
3, respectively. We note that the exact values $A_{ex},B_{ex}$
were determined with the help of the numerical computations.

\begin{center}
\scalebox{.4}{\includegraphics{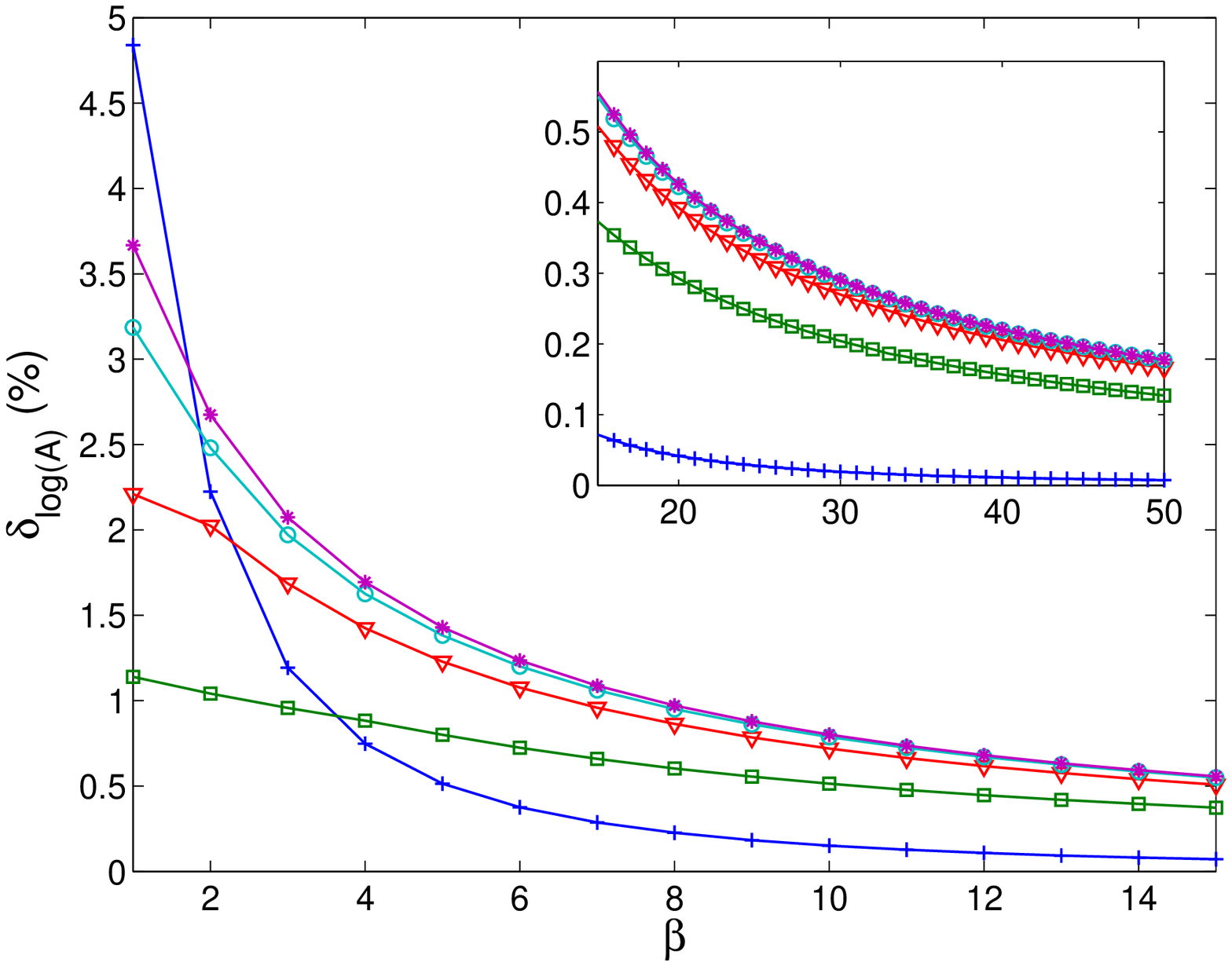}} \footnotesize

\textbf{Figure 3. Relative deviation in the approximate value of
the
normalization constant $A$ as function of parameter $\beta.$}\\

Plotted is the deviation (\ref{disc_A}) between the numerically
computed value $A_{ex}$ and the estimated value (\ref{extim_A}).
The symbols used here are consistent with the symbols in the Fig.
2.\end{center}

Considering now the tested choice for car-car potential
$V(r)=r^{-1}$ we intend to compare the equilibrium distribution
\begin{equation}
P_\beta(r)=Ae^{-\frac{\beta}{r}}e^{-Br} \label{clearance_1}
\end{equation}
with the relevant distributions of single-vehicle data measured
continuously during approximately 140 days on the Dutch two-lane
freeway A9. Macroscopic traffic density $\varrho$ was calculated
for samples of $N=50$ subsequent cars passing a detector. For the
purposes described above we divide the region of the measured
densities $\varrho \in [0,85~\mathrm{veh/km/lane}]$ into 85
equidistant subintervals and separately analyze the data from each
one of them. The sketched procedure prevents the undesired mixing
of the states with the different inverse temperature $\beta,$ i.e.
with the different density. Bumper-to-bumper distance $r_i$ among
the succeeding cars ($i$th and $(i-1)$th) is calculated (after
eliminating car-truck, truck-car, and truck-truck gaps) via
standard formula
$$r_i=v_i(t_i-t_{i-1}),$$
by means of netto time-headway $t_i-t_{i-1}$ and velocity $v_i$ of
$i$th car (both directly measured with induction-loop detector)
supposing that velocity $v_i$ remains constant between the times
$t_i,t_{i-1}$ when $i$th car and the previous one are passing a
measure point. Such a condition could be questionable, especially
in the region of small densities where the temporal gaps are too
large. However, the influence of a possible error is of marginal
importance, as apparent from the fact, that distribution function
plotted for small-density data do not show any visible deviation
from Poisson behavior expected for independent events (see Fig. 4
and relation (\ref{Poisson})).\\

We note that mean distance among the cars is re-scaled to one in
all density-regions. The thorough statistical analysis of the
traffic data leads afterwards to the excellent agreement between
clearance distribution computed from traffic data and formula
(\ref{clearance_1}) for fitted value of inverse temperature
$\beta_{fit}$ (see Fig. 4). We have obtained the fit parameter
$\beta_{fit}$ by a least-square method, i.e. minimizing the error
function $\chi^2.$ The deviation $\chi^2$ between the theoretical
and empirical clearance distributions is plotted in Fig. 5 (low
part).

\begin{center}
\scalebox{.4}{\includegraphics{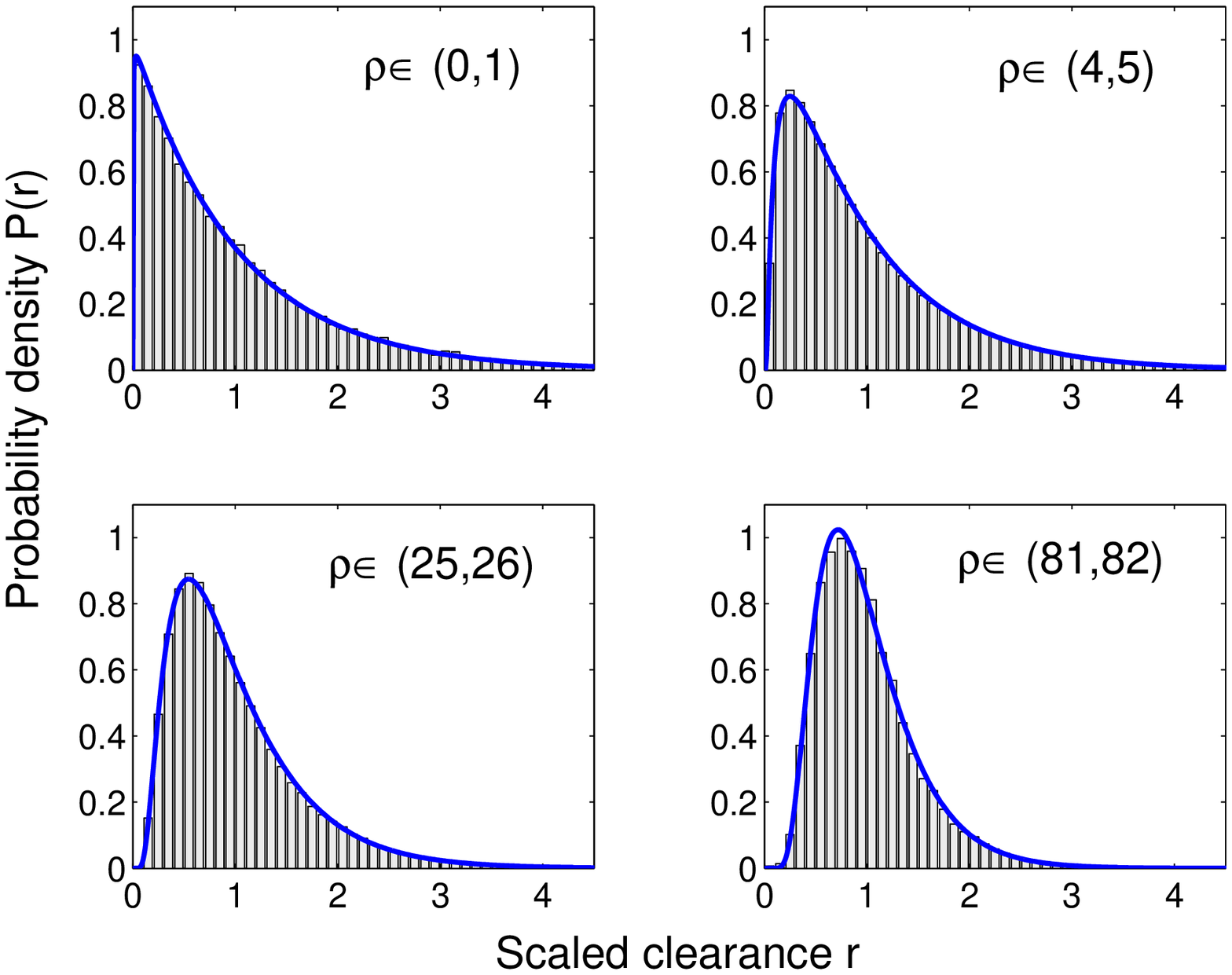}} \footnotesize

\textbf{Figure 4. Probability density $P(r)$ for scaled spacing
$r$ between successive cars in traffic flow.} \\

Histograms represent the clearance distributions computed for
traffic data from indicated density region (in
$\mathrm{veh/km/lane}$). Note that mean distance among the cars is
re-scaled to one in all density-regions. The curves represent the
predictions of statistical model (\ref{clearance_1}) for fitted
value of inverse temperature $\beta_{fit}.$ The respective values
$\beta_{fit}$ were carefully analyzed and consecutively visualized
in the following Fig. 5 (top part).

\end{center}

Dimensionless inverse temperature $\beta$ of the traffic sample,
representing a quantitative description of mental strain under
which the car-drivers are in a given situation, shows non-trivial
dependence on traffic density $\varrho$ (as visible in Fig. 5 --
top part). For free flow states $(\varrho \lesssim 20
~\mathrm{veh/km/lane})$ one can recognize a rise in temperature
having a linear behavior (up to $10 ~\mathrm{veh/km/lane}$) and
visible plateau above. In the intermediate region (between 20 and
$50 ~\mathrm{veh/km/lane}$), where free traffic converts to the
congested traffic, we detect a sharp increase in the first half.
Such a behavior can be simply elucidated by the fact that the
drivers, moving quite fast in a relatively dense traffic flow, are
under a substantial psychological pressure, which finally results
(for densities $\varrho \in [35,50]~\mathrm{veh/km/lane}$) in the
transition to the congested flows a therefore in the drop in
inverse temperature. In synchronized traffic regime ($\varrho
\gtrsim 50~\mathrm{veh/km/lane}$) the drivers vigilance rapidly
grows up which culminates by the traffic-jam formation.

\begin{center}
\scalebox{.4}{\includegraphics{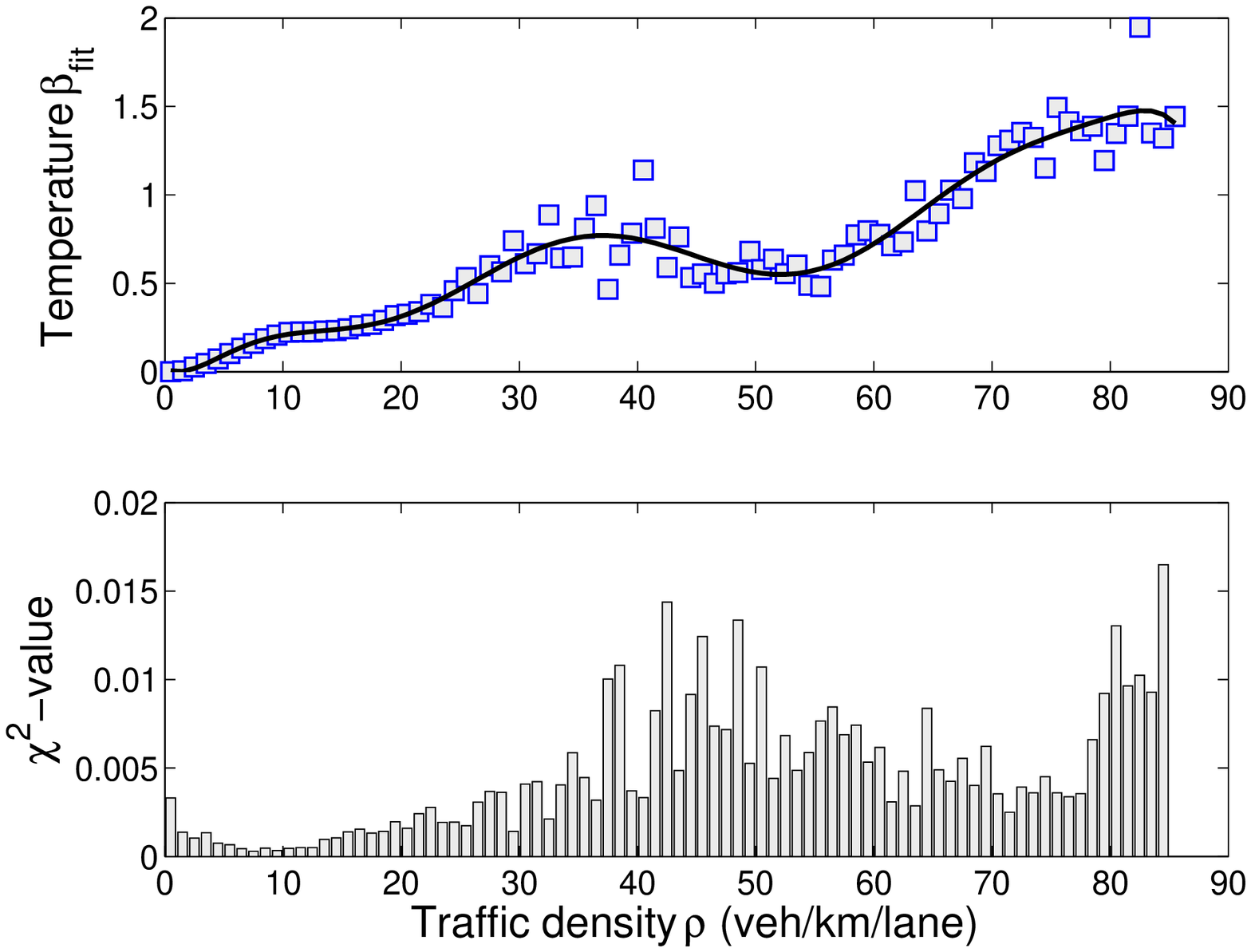}} \footnotesize

\textbf{Figure 5. Inverse temperature $\beta_{fit}$ and deviation
$\chi^2$ as a function of traffic density $\varrho.$}\\

Squares stand for values of fit parameter $\beta_{fit},$ for which
the function (\ref{clearance_1}) coincides with clearance
distribution of traffic data. The curve represents a polynomial
fit of relevant data. Bars from lower part correspond to the sums
of squared deviations between the empirical and the theoretical
netto distance distributions for $\beta_{fit}.$ We note that the
value of normalization constant $B$ was determined via formula
(\ref{main_approximation}), because the values of relevant inverse
temperatures lie in interval $[0,2],$ where the linear
approximation (\ref{extim_B}) is less suitable (see deviations in
Fig. 2). Second normalization constant $A$ was calculated by means
of equation (\ref{A_approximation}).

\end{center}

For completeness we have compared real-road clearance
distributions with probability density (\ref{spacing
distribution}) specified for power-law potentials
$V(r)=r^{-\alpha},$ where $\alpha=2,3,4,5,$ respectively. Similar
analysis was already introduced in article
\cite{Helbing_and_Krbalek}. As results from the careful analysis
of statistical deviations $\chi^2,$ relevant distributions
(\ref{spacing distribution}) -- obtained by applying a
least-square method -- predicate the inter-vehicle gaps
substantially worse than those calculated for $\alpha=1$ (follow
Fig. 6). We emphasize that small deviations $\chi^2$ (indicating a
good agreement) detected near the origin in Fig. 6 are caused by
the fact that for low traffic-densities the interactions among the
cars are vanishing and inter-vehicle gaps are therefore
practically independent. The relevant distributions in this case
came close to the Poisson distribution (see upper left-hand
subplot of Fig. 4). However, the Poisson distribution can be
obtained as a limit of distribution (\ref{LS}), i.e.
\BE \lim_{\beta \rightarrow 0_+} P_\beta(r)=e^{-r},\label{Poisson}
\EE
for arbitrary function  $V(r).$ For that reason it is impossible
to detect the interaction potential in traffic sample using the
low density data only. The comparison of various potentials
$V(r)=r^{-\alpha}$ (including the logarithmical potential
$V(r)=-\ln(r)$) brings finally the message that interaction among
the vehicles in traffic stream can be very well estimated by the
short-ranged two-body power-law potential $V(r)=r^{-1}.$ Predicted
inter-vehicle-gap distributions correspond in this case to the
computed probability density in all density intervals.

\begin{center}
\scalebox{.4}{\includegraphics{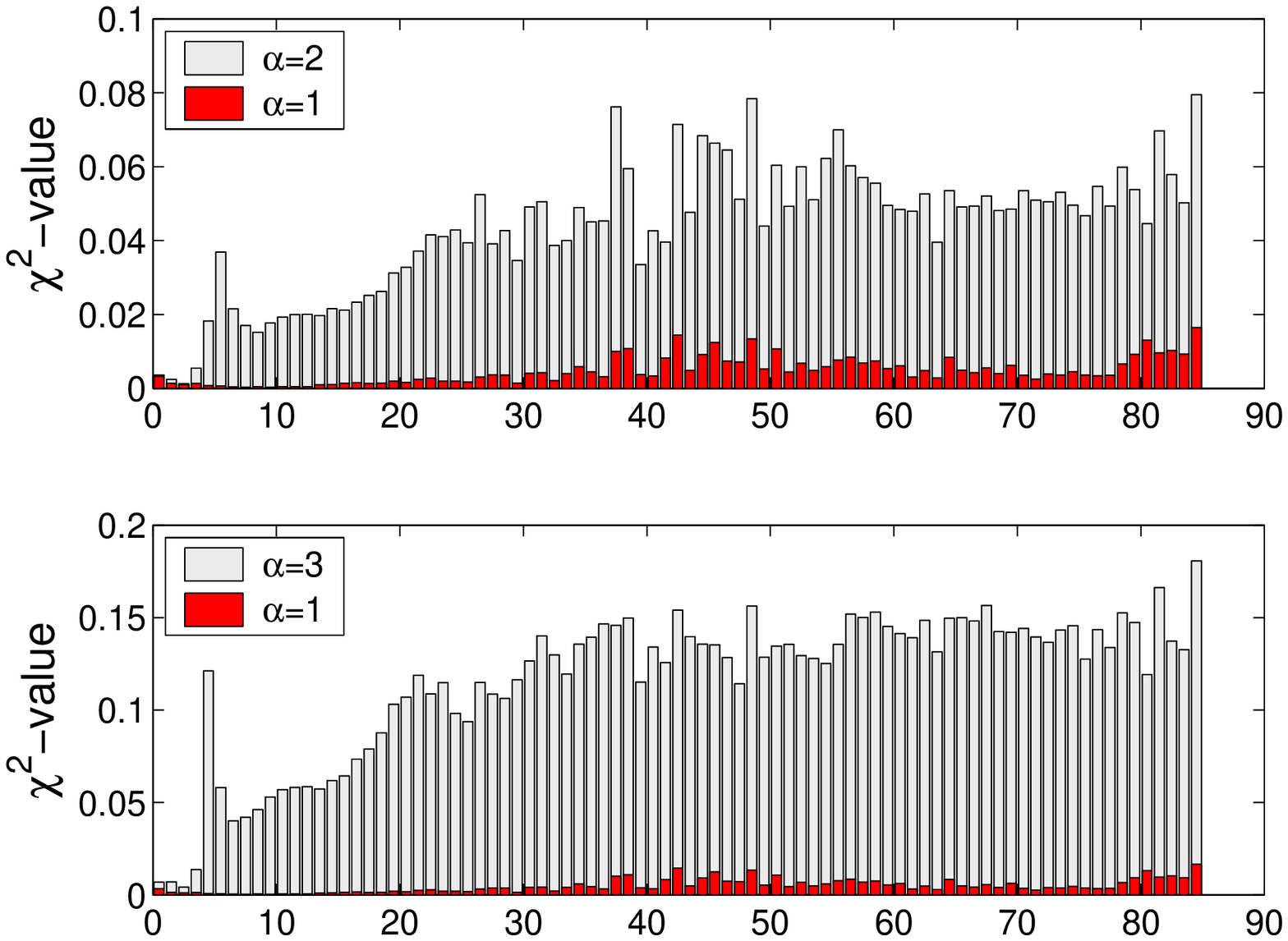}}\vspace*{0.3cm}
\\ \scalebox{.4}{\includegraphics{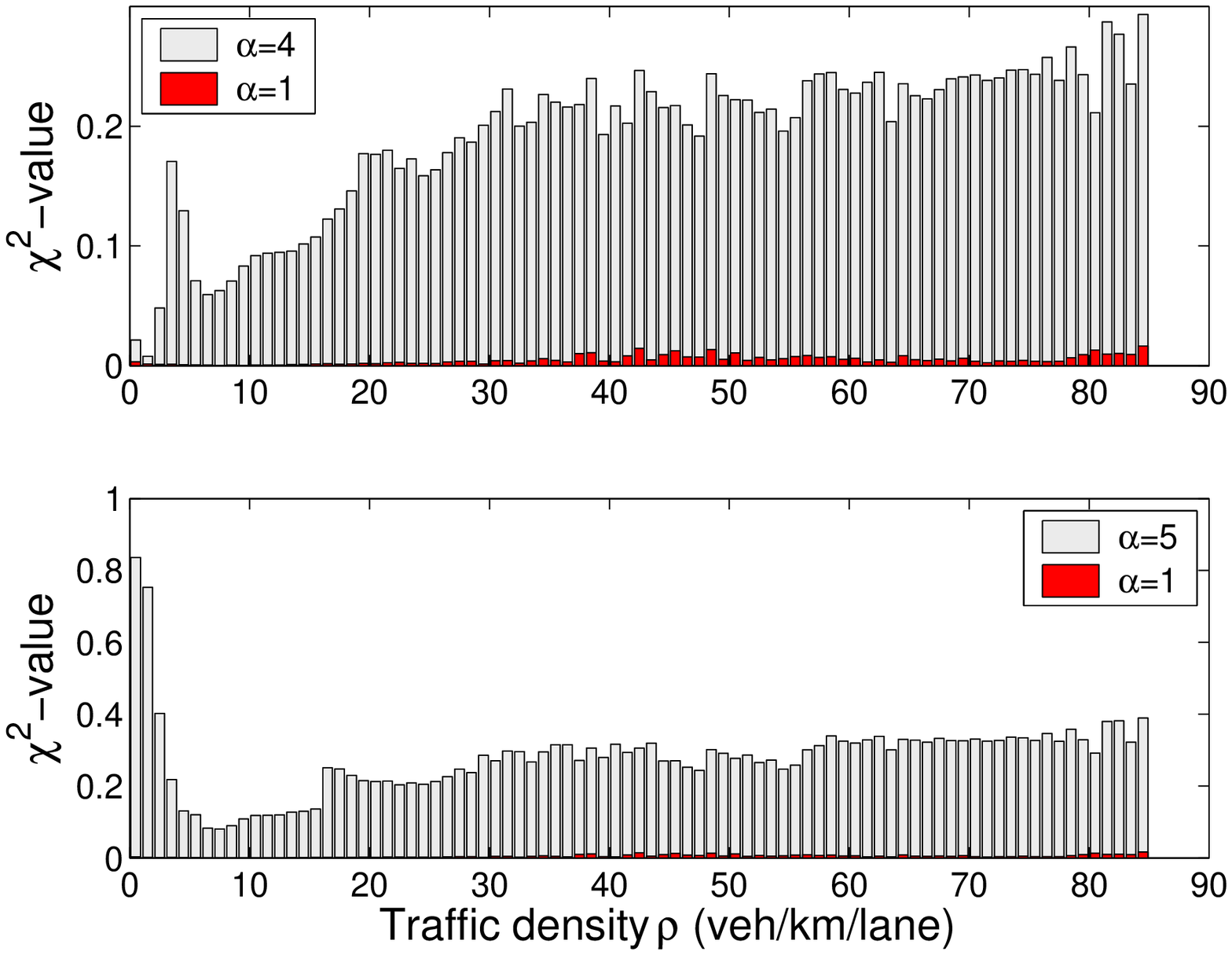}} \footnotesize

\textbf{Figure 6. Deviation $\chi^2$ between clearance
distribution of cars moving in traffic stream and normalized
distribution (\ref{spacing distribution}) evaluated for
$\alpha=2,3,4,5,$ respectively.}\\

Grey bars represent the sums of squared deviations depending on
traffic density for $\alpha=2,3,4,5,$ respectively. Dark bars
display the relevant deviations for $\alpha=1$ previously plotted
in Fig. 5 (lower part).
\end{center}

We append that another suitable quantity for comparison with
single-vehicle data is a time-clearance distribution as well.
Furthermore, time gaps among the succeeding cars are directly
measurable and therefore not burden with errors caused by
computation approximations. Determination of exact form for
time-headway probability density of above thermodynamical traffic
gas and relevant comparison with highway traffic data will be
incorporated in the continuing work. Nevertheless, several
analytical forms for corresponding distribution have already been
discussed in article \cite{HKT} and also in book \cite{May}.\\

To conclude we have found the analytical form of the
thermal-equilibrium spacing distribution for one-dimensional
traffic gas which neighboring particles are repulsed by the
two-body potential $V=r^{-\alpha},$ where $r$ is their mutual
distance. The values of two normalization constants were
successfully estimated by the convenient approximations. The
calculated distribution (for $\alpha=1$) with one free parameter
(inverse temperature $\beta$) has been compared to the distance
clearance distribution of the freeway traffic samples with the
excellent outcome. It was demonstrated that inverse temperature of
traffic sample non-trivially depends on the traffic density. The
obtained agreement between experimental and calculated
distributions confirms a convenience of the traffic potential used
for description of local traffic interactions. Presented article
crowns the quest for the mathematical formula for probability
density of mutual clearances among the cars in traffic stream and
supports the possibility for applying the equilibrium
statistical physics to the traffic modelling.\\

\emph{Acknowledgements:} We would like to thank Dutch Ministry of
Transport for providing the single-vehicle induction-loop-detector
data. This work was supported by the Ministry of Education, Youth
and Sports of the Czech Republic within the project LC06002.

\end{multicols}


\small
\begin{thebibliography}{100}

\bibitem{Bogomolny}
E.B.~Bogomolny, U.~Gerland, and C.~Schmit, Eur. Phys. J. B
\textbf{19} (2001), 121

\bibitem{Helbing}
D.~Helbing, Rev. Mod. Phys. \textbf{73} (2001), 1067

\bibitem{Helbing_and_Treiber}
D.~Helbing and M.~Treiber, preprint:
\textcolor{blue}{cond-mat/0307219}

\bibitem{Gauss}
D.~Helbing and M.~Treiber,  Phys. Rev. Lett. \textbf{81} (1998),
3042

\bibitem{HKT}
D.~Helbing, M.~Treiber, and A.~Kesting,  Physica A \textbf{363}
(2006), 62

\bibitem{Neubert}
L.~Neubert, L.~Santen, A.~Schadschneider, and M.~Schreckenberg,
Phys. Rev. E \textbf{60} (1999), 6480

\bibitem{Helbing_and_Krbalek}
M.~Krbalek and D.~ Helbing, Physica A \textbf{333} (2004), 370

\bibitem{Wagner}
M.~Krbalek, P.~Seba, and P.~Wagner, Phys. Rev. E \textbf{64}
(2001), 066119

\bibitem{Mahnke}
R.~Mahnke, J.~Hinkel, J.~Kaupu\v zs, and H.~Weber, preprint:
\textcolor{blue}{cond-mat/0606509}

\bibitem{May}
A.D.~May, \emph{Traffic flow fundamentals}, Prentise Hall,
Englewood Cliffs, New Jersey, 1990


\bibitem{Scharf}
R.~Scharf and F.M.~Izrailev, J. Phys. A: Math. Gen. \textbf{23}
(1990), 963

\end{thebibliography}
\end{document}